# Market power abuse in wholesale electricity markets

Alice Lixuan Xu [a,*], Jorge Sánchez Canales [a], Chiara Fusar Bassini [a,b], Lynn H. Kaack [a,b], Lion Hirth [a,c]

[a] *Centre for Sustainability, Hertie School, Friedrichstraße 180, 10117 Berlin, Germany*
[b] *Data Science Lab, Hertie School, Friedrichstraße 180, 10117 Berlin, Germany*
[c] *Neon Neue Energieökonomik GmbH, Karl-Marx-Platz 12, 12043 Berlin, Germany*
* Corresponding author: l.xu@hertie-school.org

**Abstract**   In wholesale electricity markets, prices fluctuate widely from hour to hour and electricity generators price-hedge their output using longer-term contracts, such as monthly base futures. Consequently, the incentives they face to drive up the power prices by reducing supply has a high hourly specificity, and because of hedging, they regularly also face an incentive to depress prices by inflating supply. In this study, we explain the dynamics between hedging and market power abuse in wholesale electricity markets and use this framework to identify market power abuse in real markets. We estimate the hourly economic incentives to deviate from competitive behavior and examine the empirical association between such incentives and observed generation patterns. Exploiting hourly variation also controls for potential estimation bias that do not correlate with economic incentives at the hourly level, such as unobserved cost factors. Using data of individual generation units in Germany in a six-year period 2019-2024, we find that in hours where it is more profitable to inflate prices, companies indeed tend to withhold capacity. We find that the probability of a generation unit being withheld increases by about 1 % per euro increase in the net profit from withholding one megawatt of capacity. The opposite is also true for hours in which companies benefit financially from lower prices, where we find units being more likely to be pushed into the market by 0.3% per euro increase in the net profit from capacity push-in. We interpret the result as empirical evidence of systematic market power abuse.

## 1. Introduction

Market power abuse is a market failure and disruptive to competitive pricing. Electricity markets are particularly vulnerable to it because electricity is largely non-storable,[1] electricity demand is rather inelastic (Hirth, Khanna, and Ruhnau 2024), and the electricity grid faces transmission constraints (Nappu, Bansal, and Saha 2013). As a result, generation companies sometimes possess large potential for temporally and spatially sensitive, unilateral market power (F. Wolak 2003; Knittel and Roberts 2005).

Capacity withholding is a classical way of exercising market power. It is the act of deliberately reducing generation output offered to the market (Twomey et al. 2005). In electricity markets consisting of single-price auctions, individual suppliers sometimes face opportunities to exercise their market power by not bidding, bidding at higher prices, or declaring a reduced amount of available capacity to unilaterally alter prices to their benefit. For example, in Germany, the Federal Cartel Office found that in 2022, the five largest generation companies had substantial opportunities to influence market prices (Bundeskartellamt 2023). In fact, scholars have attributed certain price spikes and high price periods to missing generation or "output gap," pointing to capacity withholding (Joskow and Kahn 2002; Krzywnicka and Barner 2025).

Conventional metrics on market power in terms of market share and price markups often fall short of detecting abuse in electricity markets (Harvey and Hogan 2001; Helman 2006). It is because capacity withholding does not require the generator to have a substantial market share or to withhold a large quantity to be profitable (Kwoka and Sabodash 2011). Furthermore, capacity withholding is not necessarily associated with high price levels (Bataille et al. 2019).

In this paper, we construct a method for detection of capacity withholding, by analyzing the economic incentives of companies to withhold at the hourly level. We also consider market power abuse in the opposite direction, where participants strategically push in generation (i.e. inflate output) to depress market price, as incentivized by their hedge contracts in the forward markets. We conduct a unit-level empirical analysis of suspected market power abuse in the German wholesale electricity market during the six-year period from 2019 to 2024.

We find that thermal, dispatchable units deviate in their predicted generation patterns. These deviations follow the net profit from acting strategically, either withholding profitable capacity or pushing-in cost-negative capacity. Our model estimates that the probability of a generation unit being withheld increases by about 1 % for every euro increase in the net profit from withholding one megawatt (MW) of capacity. For hours in which companies financially benefit from lower prices, we find a 0.3 % increase in the probability of a generation unit being pushed cost-negatively into the market, per euro increase in the net profit from such capacity push-in.

---

[1] Except for markets with abundant hydroelectric reservoirs and pumped storage options, most markets do not have economical options to store electricity at large scales.

The paper proceeds as follows. Section 2 provides theory and introduces our identification strategy to find potential evidence of market power abuse in electricity markets. Section 3 reviews existing empirical studies on detection of market power abuse and related literature on economic incentives and hedging. Section 4 presents the data, Section 5 the implementation of our method in three steps, and Section 6 the results of our empirical analysis. Section 7 discusses the limitations and implications of this analysis, along with relevant legal cases investigated under REMIT.[2] Section 8 concludes.

## 2. Theory

Electricity is not storable at large scales and must be consumed when it is produced. As the rather inelastic demand for electricity changes from hour to hour, the price of electricity is therefore also highly volatile, and correspondingly, generators dispatch their generation units according to the hourly varying demand. The hourly variability in the demand, price, and supply of electricity plays a crucial role in incentivizing and hence also revealing market power abuse in electricity markets.

Electricity generation companies inevitably face price risks and use forward contracts to hedge against low prices on the spot market. It is common for a company to have hedged almost all its production by the time the spot market opens one day ahead of delivery. As current forward contracts offer future products with flat production profiles over a long period of time, usually for months or a year ahead, companies can effectively hedge against price risks over the long term. Meanwhile, they are almost always under- or over-hedged in individual hours, meaning that with hourly variability in actual dispatch, it is usually the case that their production in an hour differs from the quantity they have sold forward for the overall period, either positively or negatively. When a company is under-hedged in an hour, its excess production is exposed to the spot price in that hour, allowing it to profit from a high spot price. When over-hedged, the company must pay the spot market price to meet its forward commitment, thus benefiting from a low spot price. This means that the company faces economic incentives to influence spot prices – both upwards and downwards – depending on how its hourly production differs from its forward contract.

In Figure 1, we illustrate an imaginary company's change in profit when it adjusts its hourly production to induce and benefit from a change in spot price. For both panels, we show the same fixed amount of hedged quantity $Q_j^{hedged}$ in the short term and the same supply curve for company $j$; we start from the competitive assumption that company $j$ acts as a price taker and generates at $G_{j,h}$, up to the point where its cost of production $c_j^{var}$ meets the market-clearing price $p_h^{spot}$ in hour $h$. The top panel exemplifies an hour where company $j$ happens to be under-hedged: it produces *more* than it had previously sold in forward markets ($G_{j,h} > Q_j^{hedg}$). Consequently, the company may have an incentive to withhold part of its excessive production to inflate the spot price, so as to increase its profit (in green), while bearing an opportunity cost of the missing generation (in red). Conversely, the bottom panel shows an hour where company $j$ is over-hedged: it produces *less* than it had previously sold in forward markets $(G_{j,h} < Q_j^{hedged})$. In this situation, the company is incentivized to dispatch part of its idle generation, uncalled for by the spot market, to depress the spot price, so as to save some of its hedge loss (in green), though facing a cost for the extra production and existing production at the lowered spot price (both in red). We see that in both cases, company $j$ may face an incentive to influence the spot price by deviating from its competitive generation, if the net profit from doing so is positive.

---

[2] Regulation (EU) No. 1227/2011: Regulation on Wholesale Energy Market Integrity and Transparency.

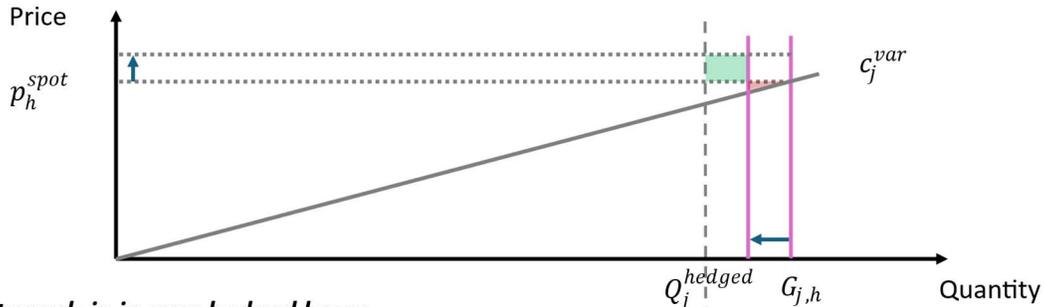

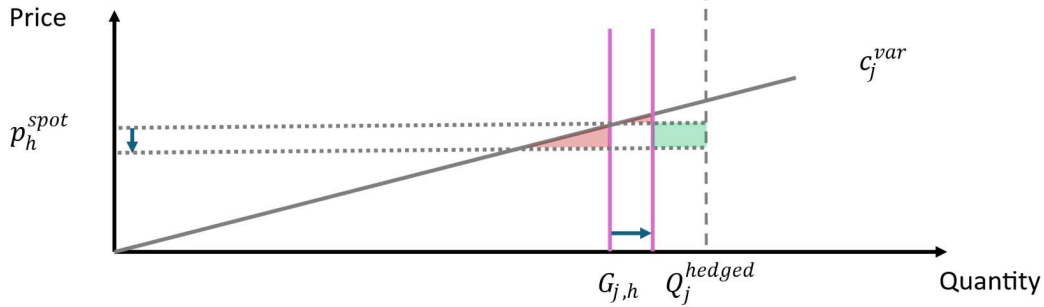

*Figure 1. Illustrative graph of economic incentives to influence spot prices for company j, considering when it is under-hedged (top panel) and over-hedged (bottom panel) in the hour. The green areas mark out the additional profit company j can earn by adjusting its generation quantities; the red areas mark out the additional cost such adjustments in generation incur for company j. This figure shows how economic incentives to exercise one's market power emerge, and provides the conceptual framework to measure such incentives, via the net profit from market power abuse.*

In this paper, we find empirical evidence of suspected market power abuse by estimating the economic incentives – to inflate prices by withholding generation capacity and to depress prices by pushing it cost-negatively into the market – and examining the observed behaviors of individual generation units. To identify uncompetitive behavior in empirical data and explain it with economic incentives, we use a three-step method. First, we identify deviation in dispatch from a constructed competitive counterfactual under price-taking assumptions, using the ex-post market-clearing price and considering start-up costs and reported outage. We then measure unit- and hour-specific indicators of the economic incentives to alter spot prices. Lastly, we examine the association between the two. The reason for taking a step-wise method, instead of a single econometric model to predict observed generation behavior, is that electricity generation technology and cost estimation is straightforward and well-understood. And with a constructed competitive counterfactual, we are able to delineate the part of observed generation behavior that is out of the ordinary, which we cannot explain given competitive assumptions. We consider the unexplained missing or excessive generation to be evidence of potential market power abuse when the association is established by an econometric model in the third step, between deviation in dispatch from the competitive benchmark and the incentives for exercising market power. This identification strategy is also consistent with suggestions for withholding analysis as a promising detection method of market manipulation (Twomey et al. 2005).

## 3. Literature

In this section, we review three strands of literature, respectively, on empirical studies of market power abuse in electricity markets, economic incentives of market power abuse, and the role of hedging. We find that most existing studies on the topic emphasize market outcomes over economic incentives and that none empirically considers hedging, though previous literature stresses their importance in electricity market power research.

Although both upward and downward price manipulations are market power abuse (Stoft 2002, 318), existing empirical studies overwhelmingly study one side – inflating prices via capacity withholding – and disregard the other – depressing prices via pushing in capacity. Previous literature sometimes classifies the withholding of generation capacity in two distinct ways:[3] financial and physical withholding (Stoft 2002, 322; Nappu, Bansal, and Saha 2013). Empirical studies using the former distinction studies the offer price of generation capacity via price-cost markups (Puller 2007) and bid patterns (Kwoka and Sabodash 2011), while the rest study missing generation output relative to competitive counterfactuals. A seminal study by Joskow and Kahn (2002)[4] finds what they call an "output gap" during an unprecedentedly high price period in California in 2000, where generation output that could be profitably sold at the competitive price was nevertheless missing on the market. Similarly, scholars have also concluded that strategic behavior may have contributed to high prices during the gas crisis in Germany in 2022 by comparing simulated outcomes from perfect competition and oligopolistic Nash-Cournot equilibrium models against empirical observations (Krzywnicka and Barner 2025).

Recent empirical analyses on capacity withholding have found positive associations between generation unit outages and increasing market prices in Sweden (Fogelberg and Lazarczyk 2014), Germany and Austria (Bergler, Heim, and Hüschelrath 2017), and Turkey (Durmaz, Acar, and Kızılkaya 2024). Though all the above studies provide consistent evidence of capacity withholding by finding an empirical relationship between market behavior and high prices, they share an emphasis on market outcomes (high prices) without explicitly modeling or explaining a key driver of the market behavior of generators: their economic incentives.

---

[3] It is pointed out that the two manifestations of withholding do not differ in strategy, but the circumstances specific to the supplier (Kwoka and Sabodash 2011).
[4] It is worth mentioning that Borenstein et al. (2002), another prominent study around the same time that also focuses on the California experience, is similar to Joskow and Kahn (2002) in methodology. It also finds evidence of market power and the exercise of it during peak demand periods, though unlike Joskow and Kahn (2002), it does not establish what the mechanism could be.

Meanwhile, scholars have pointed out that considering the generators' dynamic competitive positions and economic incentives is essential for an effective measure of market power in wholesale electricity markets (Nappu, Bansal, and Saha 2013; Bataille et al. 2019), as it can fly under the radar of conventional metrics that emphasize large players and high prices. This is because in wholesale electricity markets, any inframarginal or extramarginal generation unit can make a unilateral decision to withhold from or push in supply, to potentially achieve an inflated or depressed equilibrium price. Due to transmission constraints, unilateral market power abuse can alter prices in electricity markets even without requiring an exceptionally large market share or quantity withdrawal to become profitable, as is frequently mentioned in literature (Puller 2007; Kwoka and Sabodash 2011; Bataille et al. 2019).

From a company's perspective, the decision to exercise one's market power is contingent on one's incentive to do so: the potential profit and opportunity cost. This economic incentive is found to be associated with a company's inframarginal capacity and often, the company size. For instance, Wolfram (1998) found higher markups in empirical bids submitted for units from the larger of two generation companies in England and Wales, of similar technologies. Moreover, she finds that the incentive to set a high price for one's inframarginal capacity is moderated by the incentive to ensure that a unit is not left out of the dispatch. Her study establishes that the economic incentives that a company faces not only differ by company size and capacity mix of production types and costs, but also by the hour, where the company balances between a unit's inframarginal capacity and its likelihood to be in dispatch.

Our analysis is most comparable to Bataille et al. (2019), in that we both conceptualize the incentive for market power abuse based on the trade-off between the potential profit and opportunity cost of withholding. Bataille et al. (2019) first proposes a measure of the incentive to withhold capacity, called the Return on Withholding Capacity Index (RWC), by comparing the abuse rent gained against the lost profit margin if a supplier withheld a MW of its capacity. The authors propose that an incentive to withhold is given when the potential profit jump surpasses the threshold set by a maximally possible lost profit from the MW withheld. Though the RWC can be calculated for individual hours, the authors suggest interpretations of the RWC on an annual or half-annual basis, as more granular interpretations of the RWC are lacking without a unit- and hour-specific estimate of the opportunity cost from the capacity withheld. Nevertheless, the development of RWC constitute an important shift of focus in the literature from peak load and peak price to peak profitability that captures the economic incentives specific to the generation unit and its owner.

Existing empirical studies on market power abuse in electricity markets do not consider a company's hedge contract position, due to the lack of publicly available data on hedging. However, without this consideration, it would be wrong to determine if the company is incentivized to exercise its market power at all (F. A. Wolak 2000). For instance, Joskow and Kahn (2002) concludes with anecdotal evidence that 'the one supplier for which we do not find any significant evidence of withholding had apparently contracted most of the output of its capacity forward.' This showcases the neutralizing effect of forward markets on market power, increasing market efficiency, as established by Allaz and Vila (1993) and quantified by Ito and Reguant (2016). In fact, by simulating daily bidding strategies in competitive markets, Wolak (2000) finds significant effects of a company's hedge contract positions on its short-run incentives to raise and to reduce the market price. He points out that at sufficiently high hedge rates, a profit-maximizing generator should attempt to reduce market prices below its own marginal cost of production. Our analysis aims to address this gap in literature.

We contribute to existing literature in three ways. First, we provide a continuous measure of the hour-by-hour economic incentive to withhold or push in capacity on wholesale electricity markets. We are the first to estimate a net profit at the hourly and generation unit level that can be meaningfully interpreted across its full range. Another important novelty is that we consider price hedging and track abusive behaviors that could alter prices in both upward and downward directions. Lastly, our empirical analysis shows that those economic incentives explain the observed behavior in generation units.

## 4. Data

We obtain wholesale electricity market data from ENTSO-E Transparency Platform (ENTSO-E 2024) and unit generation and outage reports from EEX Transparency Platform (EEX 2024; Fusar Bassini 2025).[5] We then conduct manual checks to fill in the gaps for specific generation units that lack information on the two platforms. We estimate marginal cost of production for individual generation units based on estimates of their respective fuel costs (International Carbon Action Partnership 2025; Investing.com 2025a; 2025b), thermal conversion efficiency (Weibezahn et al. 2020), and start-up costs (Schill, Pahle, and Gambardella 2017, 8).[6] The preprocessed data is organized into a time series format, consisting of unit-hour observations uniquely indexed by individual generation unit's identification information (EIC code) and by the hourly time step in the six-year sample period 2019-2024.

In this analysis, we study generation from 40 thermal, dispatchable units of fuel types of lignite, hard coal, and gas that are relevant to the wholesale electricity market. We disregard units that are designated to be in capacity or network reserve, based on public information (Bundesnetzagentur 2024; 50Hertz Transmission GmbH et al. 2024). We also disregard units that serve the heat market as combined heat and power (CHP) plants because their behavior cannot be explained with data from electricity markets alone. We then exclude unit-hours reported to be unavailable.[7] Additionally, we do not study nuclear or hydroelectric units, besides accounting for company total generation where relevant. We also do not have a view on market power abuse from import generation. We assume nuclear and hydroelectric units to face dulled incentives to abuse market power and import generation to be competitive, consistent with previous studies (Borenstein, Bushnell, and Wolak 2002; Puller 2007). Lastly, there is no reason to believe that units of variable renewable energy (VRE) sources such as wind and solar cannot abuse market power, but there is limited publicly available short-run turbine- or company-level generation data. Future research is needed to study the generation behavior for any of those technologies.

---

[5] Due to limited public documentation on the preprocessing of published data on the EEX platform, we also compared data that were kindly provided to us by Bundesnetzagentur. This comparison allowed us to understand the mechanism of unit report data and the preprocessed dataset is published in (Fusar Bassini 2025).
[6] We detail the method we use to estimate marginal cost of production in Section 5.1.

The reason for focusing on wholesale day-ahead, instead of intraday or balancing markets, is that the day-ahead market is the largest and most relevant for market power abuse. In Germany, the day-ahead electricity market accounts for about two-thirds of all trade volume in 2024, while intraday and balancing markets are not auction-based and rather small (EPEX SPOT 2025). Additionally, potential abuse in the German balancing market is found to be modest (Just and Weber 2015) and considered second-order to that in the day-ahead market, particularly in the short run (Duso, Szücs, and Böckers 2020).

## 5. Method

To identify market power abuse, we employ a three-step methodology. We first use an optimization model to estimate the profit-maximizing dispatch of generation units, assuming competitive (price-taking) behavior. This gives us a time series of competitive dispatch estimates for each generation unit, which we then compare against observed generation behavior, to identify deviation from competitive generation. As a second step, we measure the economic incentives for market power abuse by calculating the net profit from capacity withholding and push-in, accounting for output and input prices as well as forward hedging. Thirdly, we identify the association between deviation from competitive generation and the economic incentives to do so by regressing the step-one output against the step-two output. The econometric approach in the third step allows us to identify market power abuse by distinguishing any deviation in generation that is driven by economic incentives from the rest that results from any error in estimating competitive dispatch.

In the remainder of this section, we outline the three steps. Figure 2 gives an overview of our method and illustrates the three steps we take in our analysis.

---

[7] Outage can be reported for strategic reasons, instead of genuine, technical problems. Previous literature suggests that this happens in Germany (Bergler, Heim, and Hüschelrath 2017), but our current analysis does not include these types of market power abuse because the extent of such illegitimacy of outage reports is unclear. For this reason, a recent study on strategic generation behavior also exclude strategy via outage (Krzywnicka and Barner 2025).

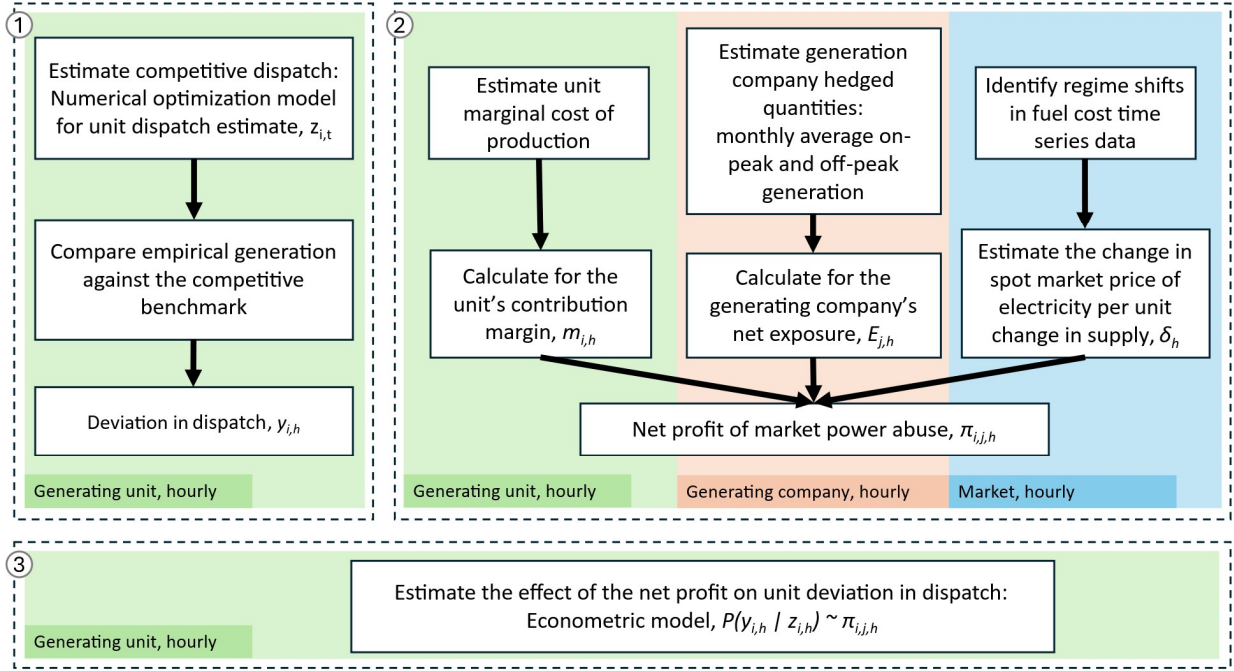

Figure 2. Conceptual map of our three-step method in this analysis. We outline the respective processes to estimate the outcome (y) in step 1 and predictor variable (π) in step 2 that serve as inputs for the regime-switching logit regression model in step 3.

## 5.1 Competitive dispatch

As a first step, we determine the profit-maximizing hour-by-hour dispatch of each individual generation unit using a numerical optimization model. This way, we can model the dispatch decisions while considering inter-temporal constraints, reflecting the set of conditions that informs the final dispatch decisions: spot prices, thermal conversion efficiency, variable and start-up costs, minimal load.[8] The goal is to have a benchmark of the hourly dispatch of each generation unit under perfect competition, for which we assume price-taking behavior. We determine this benchmark by solving a mixed-integer linear optimization problem that maximizes short-term profit, independently for each individual generation unit. The profit equation that the model seeks to maximize is:

$$\pi_{i,H} = \sum_{h \in H} (p_h * G_{i,h} - c_{i,h}^{var} * G_{i,h} - c_{i,h}^{start} * s_{i,h})$$

Here, revenues correspond to the income made by selling electricity generated $G_{i,h}$ at the market price $p_h$. The costs of operating the power plant are the variable cost of producing electricity $c_{i,h}^{var}$, derived from the prices of fuel and carbon emissions and the thermal efficiency of the plant, and the startup costs of turning the plant on $c_{i,h}^{start}$, which include additional fuel requirements and plant depreciation, depending on the status of the plant, $s_{i,h}$. Assumptions on the technology specific minimum running load and startup costs follow Schill, Pahle, and Gambardella (2017).

---

[8] A detailed formulation of this model, including all technical constraints, is included in Appendix A, as well as more information on the input parameters, the decision variables and our approach for solving the problem for all units in our sample.

For each generation unit, we run the competitive dispatch model at an hourly granularity under perfect-foresight assumptions within a horizon of 5 weeks. To allow the model sufficient information for each end of the horizon, we use rolling windows with overlapping horizons, where the window is rolled forward every 4 weeks after each horizon of 5 weeks.

We implement a Monte Carlo approach with $N = 1000$ iterations to account for uncertainty around input parameters. Specifically, we vary the thermal conversion efficiency, the fuel requirement for a cold start, and the local fuel costs of each power plant, key parameters that impact our estimates for both variable and startup costs.

Using the objective function and the set of constraints detailed in Appendix A, the model decides the optimal value for the outcome variables. The three outcome variables of the model are the unit status $\hat{d}_{i,h}$ (whether the plant is on or off), the level of the unit's output $G_{i,h}$ (a continuous variable in MW), and the startup decision for that hour $s_{i,h}$ (whether the unit was started in that hour or not). The key output of the model is the power plant status, which we denote $\hat{d}_{i,h}$. Its value can be summarized as:

$$\hat{d}_{i,h} = 0 \text{ if } G_{i,h} = 0 \text{ and } \hat{d}_{i,h} = 1 \text{ otherwise}$$

We then aggregate the binary output $\hat{d}_{i,h}$ (the on/off state of the plant) of each model run into a continuous index $\bar{d}_{i,t}$ ranging from 0 to 1, such that

$$\bar{d}_{i,h} = \frac{1}{N} \sum_{n=1}^{N} \hat{d}_{i,h}^n, \qquad \bar{d}_{i,h} \in [0,1]$$

We interpret the average across model runs $\bar{d}_{i,h}$ as the likelihood of unit $i$ at hour $h$ to be in dispatch under competitive assumptions. To increase confidence in our analysis results, we discard hours where the estimated status $\bar{d}_{i,h}$ is sensitive to the input parameters of the model and keep only those hours where the simulation results are consistent for at least 95% of the runs. That means, we keep results for $\bar{d}_{i,h} \leq 0.05$ and $\bar{d}_{i,h} \geq 0.95$ and code this information in variable $z_{i,h}$, denoting the estimated competitive counterfactual.

$$z_{i,h} = \begin{cases} 1, & \bar{d}_{i,h} \geq 0.95 \\ 0, & \bar{d}_{i,h} \leq 0.05 \end{cases}$$

Correspondingly, we code empirical dispatch $d_{i,h}$ as a binary status variable, where 1 denotes dispatch and 0 denotes non-dispatch, and compare it against $z_{i,h}$. This gives us unit deviation in dispatch from competitive dispatch, i.e., the difference between $d_{i,h}$ and $z_{i,h}$, which we denote $y_{i,h}$. The resulted deviations in dispatch are thus in three discrete categories: significant deviation in the negative direction (missing or undersupplied generation), no significant deviation, and significant deviation in the positive direction (surplus or oversupplied generation), with respective discretized values, such that

$$y_{i,h} = \begin{cases} 1, & if\ d_{i,h} = 1\ and\ z_{i,h} = 0 \\ 0, & if\ d_{i,h} = z_{i,h} \\ -1, & if\ d_{i,h} = 0\ and\ z_{i,h} = 1 \end{cases}$$

The main reason for discretizing dispatch estimates is that from an economic standpoint, it is never optimal for a power plant to generate between minimum and maximum capacity constraints. When it is profitable to generate, it should generate at its maximal capacity; when it is not profitable to generate but the loss is preferable over the startup costs, it should keep generation at the minimal load level; when it is not profitable and the loss outweighs the costs of a startup, it should shut down. We sometimes nevertheless find power plants generating at levels in between the minimum and maximum capacity constraints due to technical reasons such as ramping and mistakes on the operator side, and by discretization, we effectively exclude the influence from those aspects on the variable and focus on the binary distinction of the generation decision.

The result of this first step in our methodology, $y_{i,h}$, is the hourly time series of the deviation of each generation unit from its estimated competitive behavior, i.e. negative deviation in dispatch, and positive deviation in dispatch, or no significant deviation. Overall, we find that the observed generation does not significantly differ from our estimates for a competitive dispatch for 83 % of the time. We have identified negative deviation in dispatch in 8 % and positive deviation in dispatch in 10 % of the unit-hours in the sample.

## 5.2 Economic incentives for market power abuse

As a second step, we estimate, for each hour $h$ and for each generation unit $i$, the economic incentive to inflate or depress the market price by withholding or pushing in capacity. The incentive is determined by three factors:

- The slope of market supply ($\delta_h$): With a steeper supply curve, altering dispatch has a larger impact on spot prices.
- The net exposure of the company ($E_{j,h}$): When the company faces a larger difference between the quantity it is generating ($G_{j,h}$) and that it has sold on the forward market ($Q_j^{hedged}$), changes on spot prices have a larger impact on its profit.
- The contribution margin of the generation unit ($m_{i,h}$): A larger positive (negative) difference between the spot price and its variable cost of production means a higher profit (cost) to dispatch and hence a higher opportunity cost to withhold (push in) the unit.

The first factor is a property of the whole market, the second a property of the generation company, and the third a property of the generation unit. All three properties change from hour to hour. We find that together, they measure the net profit from marginal changes in generation. We derive net profit from the company profit equation (Equation 1), where the generation $G_{j,h}$, of company $j$ in hour $h$ receives the spot price $p_h^{spot}$ at a variable cost of production $c_h^{var}$, a pre-determined quantity of which is hedged at a forward price $p_h^{forward}$ in exchange of the spot price.[9]

$$\pi_{j,h} = G_{j,h} * (p_h^{spot} - c_h^{var}) + Q_{j,m}^{hedged} * (p_h^{forward} - p_h^{spot}) \qquad (1)$$

The marginal change in profit $\pi_{j,h}$ for company $j$, when it considers changing generation using its generation unit $i$ in a certain hour, is thereby calculated as the first partial derivative of $\pi_{j,h}$, $\pi_{i,j,h}'$ by the generation unit $i$ and by the direction of the change in generation (negative for withholding and positive for push-in). Equation 4 and 5 denote the marginal change in profit, i.e. the net profit from potential market power abuse via capacity withholding and push-in.

$$G_{j,h} = \sum G_{i,h}, \ \forall i \in j \qquad (2)$$

$$\pi'_{i,j,h} = \frac{d\pi_{j,h}}{dG_{i,h}} \qquad (3)$$

$$\pi_{i,j,h}^W = \delta_h * (G_{j,h} - Q_{j,m}^{hedged}) - (p_h^{spot} - c_{i,h}^{var}) \qquad (4)$$

$$\pi_{i,j,h}^P = -\delta_h * (G_{j,h} - Q_{j,m}^{hedg}) + (p_h^{spot} - c_{i,h}^{var}) \qquad (5)$$

where $\pi$ stands for profit, $p$ for price, $c$ for cost of production, $\delta$ for marginal change in spot price, $G$ for generation quantity and $Q^{hed}$ generation quantity committed to the company's forward contracts, $i$ for generation unit index, $j$ for company index, $h$ for hourly time-step index and $m$ for monthly time-step index.

The rest of this subsection details our operationalization of the three factors we measure to calculate the net profit from market power abuse ($\pi_{i,j,h}^{W|P}$) via withholding or pushing in individual units.

The first factor in determining the net profit is the slope of the aggregate supply curve. The slope determines how much the spot price will respond to an additional MWh of generation. Specifically, the slope that we refer to is the change in the spot price (in EUR per MWh) per megawatt (MW) of capacity either withheld from or added to the market. This slope must be estimated empirically for all hours in our sample.

---

[9] Start-up cost is not in the profit equation here because it does not affect the derivation of change in profit by marginal change in generation. It is, however, always a part of the cost considerations when a power plant shuts down or starts up. We consider start-up costs and minimum load constraints to identify deviation in dispatch, but not in our marginal net profit estimates. We explain the reasoning for this specification further in Section 6.3.

The shape of the supply curve at a point in time is determined by available installed capacity, ranked by variable costs. Thermal and hydro units are generally dispatchable unless they undergo planned or forced outages, while VRE sources depend on weather conditions. Thus, the three primary drivers (excluding market power) of the supply curve's shape are[10]: 1) capacity, which is relatively stable over time; 2) weather, which influences VRE availability; 3) fuel costs, which significantly affect variable costs and are especially volatile in our sample period due to the energy crisis. First, to control weather-driven VRE fluctuations, we use residual load, defined as total demand minus renewable generation. Since VREs have near-zero marginal costs, their main effect is to shift the supply curve horizontally without altering its slope. Second, to capture changes in fuel costs, we cluster the time series based on fuel price regimes. We formulate the slope of the supply curve we estimate as:

$$\delta_h = \frac{\partial \widehat{p_h}}{\partial l_h} = \frac{\partial f(l_h, C_h)}{\partial l_h} \qquad (6)$$

Here, $p$ denotes price, $l$ residual load, $C$ a set of fuel cost components that consists of price indices for natural gas, hard coal, and allowance for carbon emission under the European Union Emissions Trading Scheme (EU ETS).

To identify shifts in the thermal supply curve's structure, we cluster time periods based on the dynamics of coal and gas prices, both adjusted for their carbon emission costs. We use these two time series jointly because coal and gas-fired power plants are typically the price-setting technologies in Germany. We apply a dynamic programming algorithm[11] to detect structural breakpoints in the fuel price series. The number of breakpoints is increased incrementally until we reach 11,[12] at which point the clustering explains 95% of the variance in the series (measured as the reduction in the sum of squared errors relative to a model with no breakpoints). Each cluster thus corresponds to a regime where the relationship between gas and coal prices is relatively stable. Figure 3 presents the empirically identified regime shifts in electricity supply over the sample period.

---

[10] For papers that discuss the effect of supply-side drivers on electricity prices see, for example, (Abrell, Kunz, and Weigt 2008; Manera, Serati, and Plotegher 2008; Chen and Bunn 2010; He et al. 2013; Maciejowska 2020; Shah, Iftikhar, and Ali 2022)

[11] We use an algorithm similar to K-means but which enforces clusters to be contiguous. For a given number of clusters, it identifies the breakpoints in the series that lead to the clusters that minimize variance. Each cluster is defined by its centroid, a pair of average coal and gas prices. The algorithm seeks to minimize the sum of total within-cluster squared errors.

[12] We have tested results for the number of breakpoints up to 20 and find that changing it has a negligible effect on final model results. We choose a lower number of breakpoints because our goal is not to simply minimize variance, but to find points that, at the time, could be perceived by market participants as constituting a regime shift. Therefore, we prefer not to overfit ex-post.

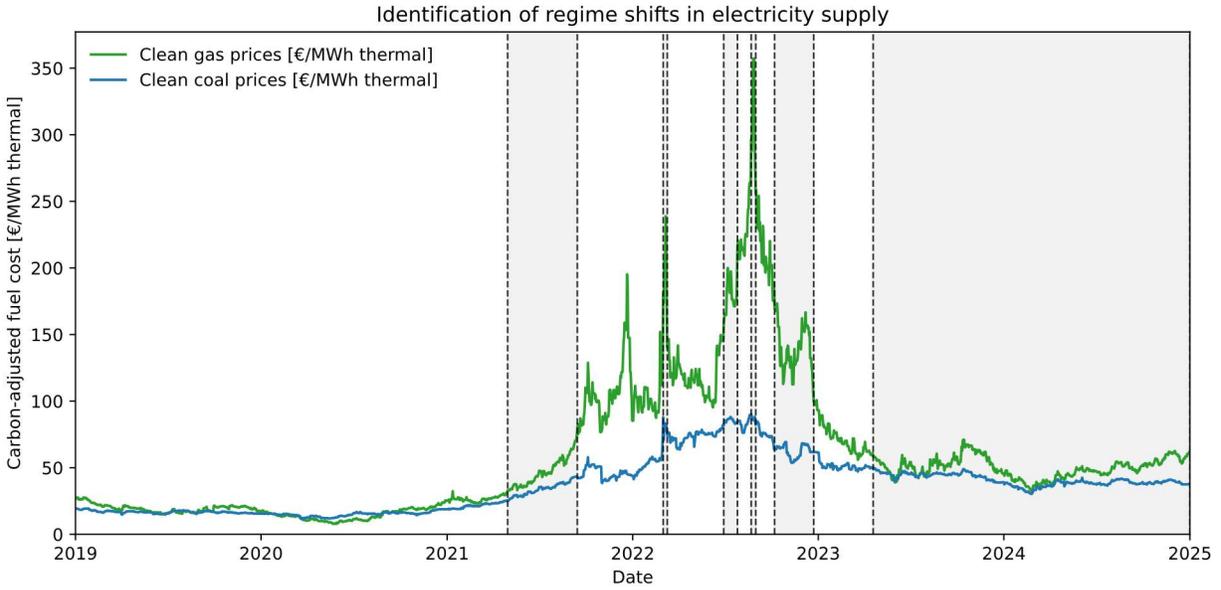

*Figure 3. Time series of carbon-adjusted gas and coal prices over the years 2019-2024, where vertical dashed lines mark regime shifts in electricity supply, as identified by the clustering algorithm.*

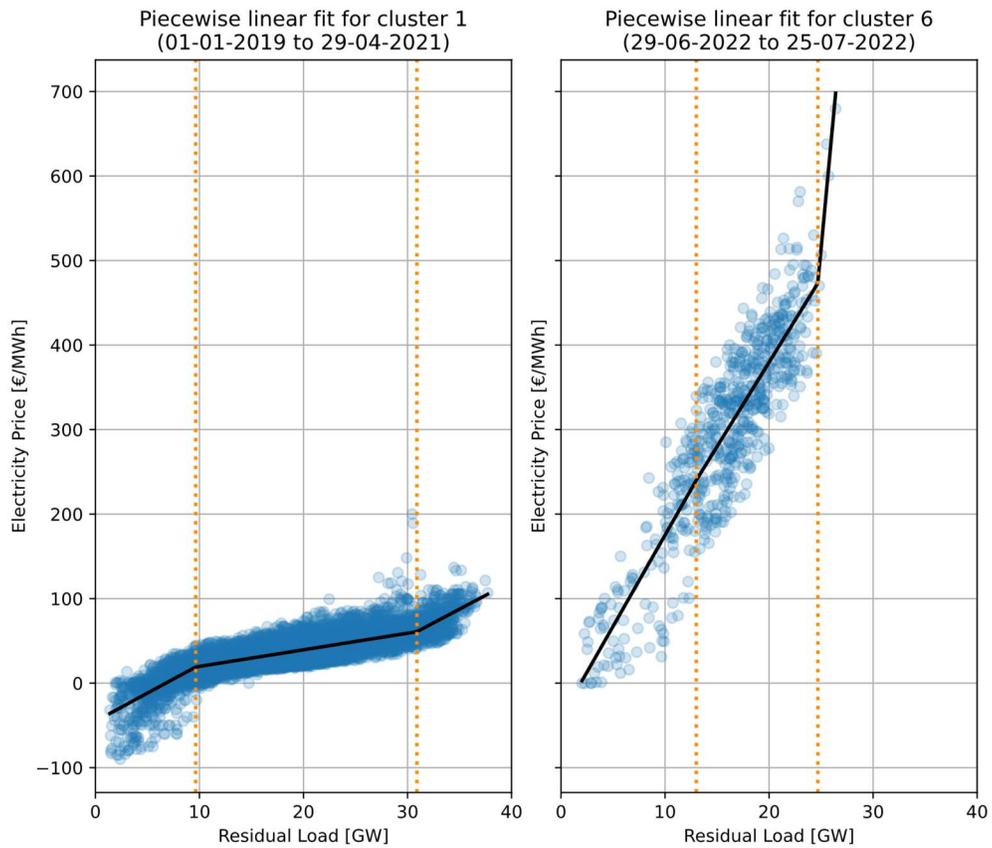

*Figure 4. Estimated supply curve within each regime, marked in black, based on a piecewise linear fit for each cluster. Two of the twelve clusters are presented here, as an example. The orange dotted vertical lines mark when changes in the slope of the supply curve is detected along residual load.*

Within each identified fuel prices regime, we then estimate the supply curve's slope using data-driven, piecewise linear fits between electricity price and residual load. This approach reflects the typical structure of electricity supply, which is often modeled as a step function or piecewise linear function (Stoft 2002, 225). This allows us to capture the marginal increase in prices as residual load increases within each fuel cost regime.

The second factor in determining the net profit is net exposure, estimated as the difference between company total generation and hedged quantities in a given hour. We estimate company hedged quantities with its average total generation output respectively for on-peak and off-peak hours[13] of each month. We assume that companies optimize their hedge contracts with perfect foresight within the monthly horizon and leverage two baseload futures products – monthly on-peak and off-peak – to hedge against short-term price risks. Net exposure is as expressed below.

$$E_{j,h} = G_{j,h} - Q_{j,m}^{hedged} \qquad (7)$$

$$Q_{j,m}^{hedged} = r * \bar{G}_{j,m} = r * \frac{1}{H}\sum_{h=0}^{H} G_{j,h} \qquad (8)$$

where $E$ stands for net exposure, $G$ for generation, $r$ for hedge rate, $h$ for the hourly index, and $H$ for the number of hours in the month. Figure 5 shows company generation and estimated net exposure at the hourly granularity, for the calendar month of April 2024.

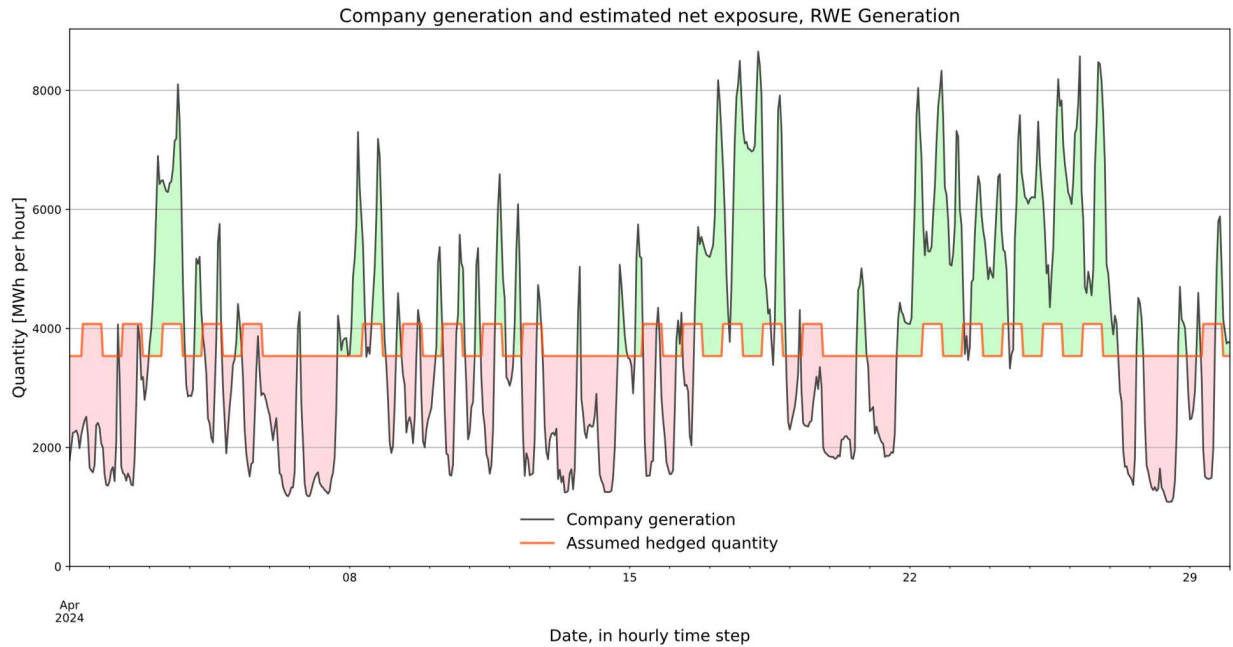

*Figure 5. Company generation (physical position in black line) against assumed hedged quantity (forward position in orange line), varying by on-peak (8 am to 8 pm every weekday) and off-peak hours. The difference between the two gives the company's estimated*

---

[13] On-peak hours include 12 hours between 08:00 and 20:00 every weekday and off-peak hours include all other hours, as is standard in European electricity markets.

*net exposure, shaded in green when positive and in red when negative. The example time frame is the calendar month of April 2024, and the example company is RWE Generation.*

Effectively, previous empirical studies that assume away hedging assume a hedge rate of zero and hence disregard any potential incentives to push-in capacity. We assume the hedge rate to be 1 for our main model specification, as we believe it to be the most realistic assumption based on publicly available information from German generation companies. Results of sensitivity runs with assumed hedge rates at lower levels are also included in Section 6.3.

The third and last factor we estimate is the contribution margin, captured by the EUR per MWh difference between spot price and unit cost of production, should the unit in question be withheld or pushed in. We estimate the unit's contribution margin to indicate the unit-level forgone profit, expressed as

$$m_{i,h} = p_h - c_{i,h} \tag{9}$$

After operationalizing the three components, we use Equations 4 and 5 to calculate net profit. We find that the net profit of withholding or pushing in one MW generation capacity is generally negative, where withholding is profitable for 18 % and push-in for only 4 % of the sample unit-hours. Meanwhile, the range of net profit is extensive, with a distribution very much skewed to the left, as shown in Table 1. This suggests that capacity withholding and push-in are seldom profitable, but when conditions are favorable in rare cases, both withholding and push-in can be incredibly profitable.

|  | *Minimum* | *50 % quantile* | *90 % quantile* | *99 % quantile* | *Maximum* |
|---|---|---|---|---|---|
| *Net profit* | | | | | |
| *From capacity withholding [EUR/MW]* | -875 | -13 | 4 | 32 | 197 |
| *From capacity push-in [EUR/MW]* | -985 | -34 | -6 | 10 | 175 |

*Table 1. Summary statistics of estimated net profit from capacity withholding and push-in.*

Taking a close look at the net profit from market power abuse at the hourly level, we see that it has high hourly variability, driven by the hourly variation in the supply of and demand for electricity. Though rare and to small extents, a lignite unit can sometimes be profitable withheld, and a gas unit can sometimes be profitable pushed in, despite their margins causing a large opportunity cost for their deviation in dispatch. Figure 6 gives an example of four weeks in March 2024 for an example gas-fired unit of Uniper.

Furthermore, for a given unit, the opportunity to withhold only exists when the unit is predicted to be in dispatch in the first place, and similarly, the opportunity to push in a unit only exists when the unit is out of dispatch under the assumptions of perfect competition. That is a prerequisite for any net profit estimate of potential withholding and push-in to give meaningful measurement of incentives. To detect market power abuse in empirical generation, it therefore only makes sense to focus on unit-hours where the opportunities to exercise one's market power exist, not where supposed incentives are inactionable. We explain how we implement this step in our econometric model setup in Section 5.3. Figure 6 also shows the estimated net profit from market power abuse when the opportunity to act on it exists and when it does not.

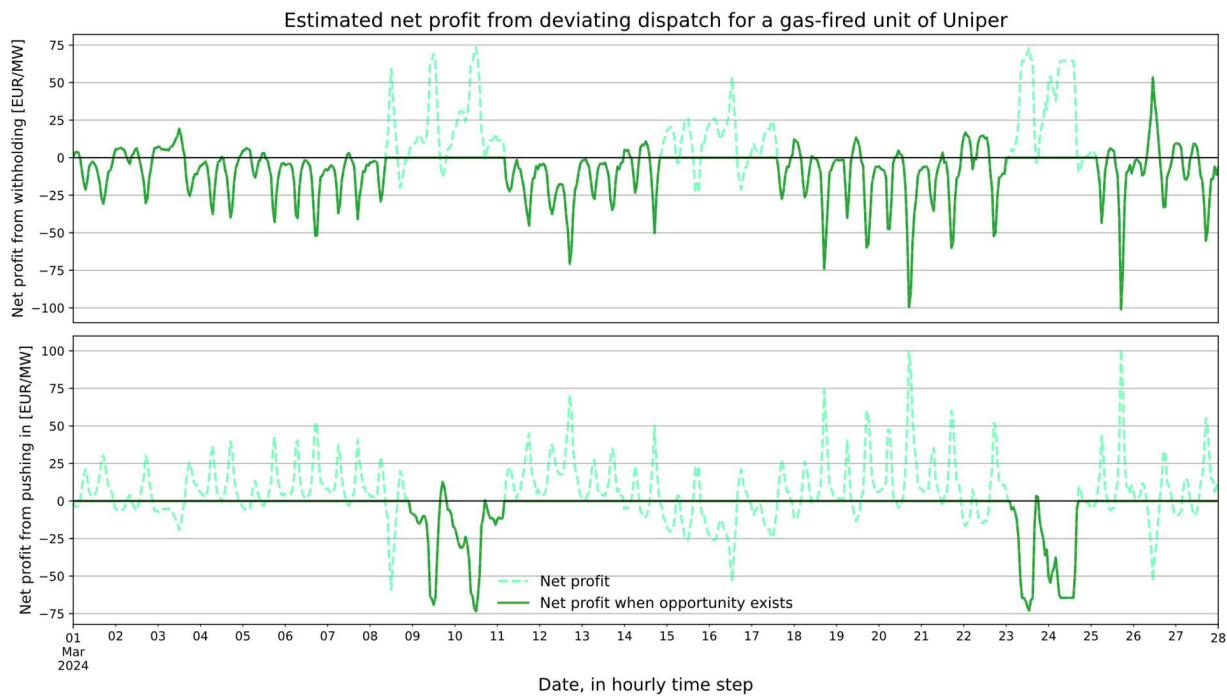

*Figure 6. Hourly net profit from withholding capacity (in top panel) and that from pushing in capacity (in bottom panel), over a sample period of four weeks in March 2024, for a sample gas-fired unit owned by Uniper.*

## 5.3  Econometric approach

In the third and last step of our method, we correlate deviation in dispatch $y_{i,h}$ (as estimated in step 1) with the economic incentives for doing so $\pi'_{j,h}$ (as determined in step 2). We hypothesize that generators tend to withhold capacity more often if it is more profitable for them to do so and, vice versa, push capacity into the market more frequently if that is more profitable. If part of the observed deviation in dispatch $y_{i,h}$ is the result of intentional, strategic behavior to abuse market power, then there will be an effect from $\pi'_{j,h}$ on $y_{i,h}$.

We apply a logit regression model to predict the probabilities of deviation in dispatch, in either direction, from the competitive benchmark, along variations of the net profit from market power abuse. Because the generation outcome is structurally restricted by competitive dispatch estimates, meaning that a negative deviation in dispatch is only possible when the competitive benchmark estimates dispatch and vice versa, the switch between having either deviation option available can be seen as a regime shift that is determined by the competitive benchmark outside the econometric model. We apply an exogenous regime-switching logit model, using maximum likelihood, to predict deviation in dispatch against no deviation in two separate regimes: when dispatch is estimated under competitive assumptions ($z_{i,h} = 1$) and when it is not ($z_{i,h} = 0$).

The probability of unit deviation in dispatch $y_{i,t}$, conditional on the range of $\bar{d}_{i,t}$ that determines the regime, such that

$$P(y_{i,h} = -1 | \pi^W_{i,j,h}, z_{i,h} = 1) = \frac{\exp(\beta_0^{z_{i,h}} + \beta_1^{z_{i,h}} \pi^W_{i,j,h})}{1 + \exp(\beta_0^{z_{i,h}} + \beta_1^{z_{i,h}} \pi^W_{i,j,h})} \tag{10}$$

$$P(y_{i,h} = 1 | \pi^P_{i,j,h}, z_{i,h} = -1) = \frac{\exp(\beta_0^{z_{i,h}} + \beta_1^{z_{i,h}} \pi^P_{i,j,h})}{1 + \exp(\beta_0^{z_{i,h}} + \beta_1^{z_{i,h}} \pi^P_{i,j,h})} \tag{11}$$

where $y_{i,h}$ denotes unit deviation in dispatch, $z_{i,h}$ denotes the regime, and $\pi^{W|P}_{i,j,h}$ denotes the net profit from market power abuse, for unit $i$ in hour $h$.

We believe the model, after controlling for predicted dispatch $z_{i,h}$, does not need additional control variables to allow for a causal interpretation.[14] We believe that there is no substantial reason to worry about omitted variable bias here because there is no other association between the outcome variable Y (deviation in dispatch) and predictor variable X (net profit from market power abuse) than what is already captured by the model. For example, both deviation in dispatch and the economic incentives depend on price levels, but after conditioning on predicted status under competitive assumptions (which contains the relevant information about prices), we find no further causal mechanism between price and Y that does not happen via X.

## 6. Results

In this section, we present our findings on the overall market, followed by results from year-, technology- and company-specific model runs. We also include model sensitivity results by varying levels of hedge rate and changing the unit of account from per-megawatt to per-unit.

### 6.1 Main results

Our main specification consists of two mutually exclusive logit models. The first regresses unit-hour observations of suspected capacity withholding (where we see no generation despite our competitive benchmark suggesting it is profitable to run) against our unit-hour-specific estimates of how profitable it is to withhold capacity. The second model examines suspected cases of capacity push-in (where we see generation despite our competitive benchmark suggesting it is not profitable to run) in association with our estimate of the profitability to push in that unit capacity. We find that our competitive dispatch estimates match empirical generation behavior for 82.5 % of the sample. The residual differences between empirical observation and competitive counterfactual lie in the negative direction for 8.0 % and in the positive direction for 9.5 % of the sample. In the following, we discuss the associations we find between when deviation in dispatch is observed and when the potential net profit from such deviation is higher, when most of the generation behavior is already accounted for by competitive counterfactuals.

---

[14] The model can be improved by including demand or drivers of demand such as temperature. However, demand is very much inelastic and hence its downstream effect on market-power-induced price change is expected to be minimal. Meanwhile, adding variables risks introducing spurious associations between the predictor and the outcome variables.

As we are using a logit model to predict exceptional events that are relatively rare and nuanced, the model's goodness of fit is expected to be low and correspond to the general competitiveness of the market. For perfectly competitive markets, for instance, we do not expect deviation in dispatch to have any strategic component that can be predicted by the net profit from market power abuse. In Table 2, we show the logit regression results, as well as the pseudo $R^2$, a ratio index consisting of the log-likelihood of the model with the predictor and that of the model with only the intercept (McFadden 1974).

|  | *Negative deviation in dispatch (suspected withholding)* | *Positive deviation in dispatch (suspected push-in)* |
|---|---|---|
| ***Intercept*** | -1.1660*** | -0.9327*** |
| ***Net profit from capacity withholding [EUR/MW]*** | 0.0102*** |  |
| ***Net profit from capacity push-in [EUR/MW]*** |  | 0.0034*** |
| ***$R^2$ (McFadden)*** | 3.1 % | 0.6 % |
| ***Number of observations*** | 727741 | 686341 |

*Table 2. Logit regression results for two models: one for suspected capacity withholding among negative deviation in dispatch, the other for suspected capacity push-in among positive deviation in dispatch. Note: \* signifies statistical significance at p-value of 0.05, \*\* at p-value of 0.01, \*\*\* at p-value of or below 0.001.*

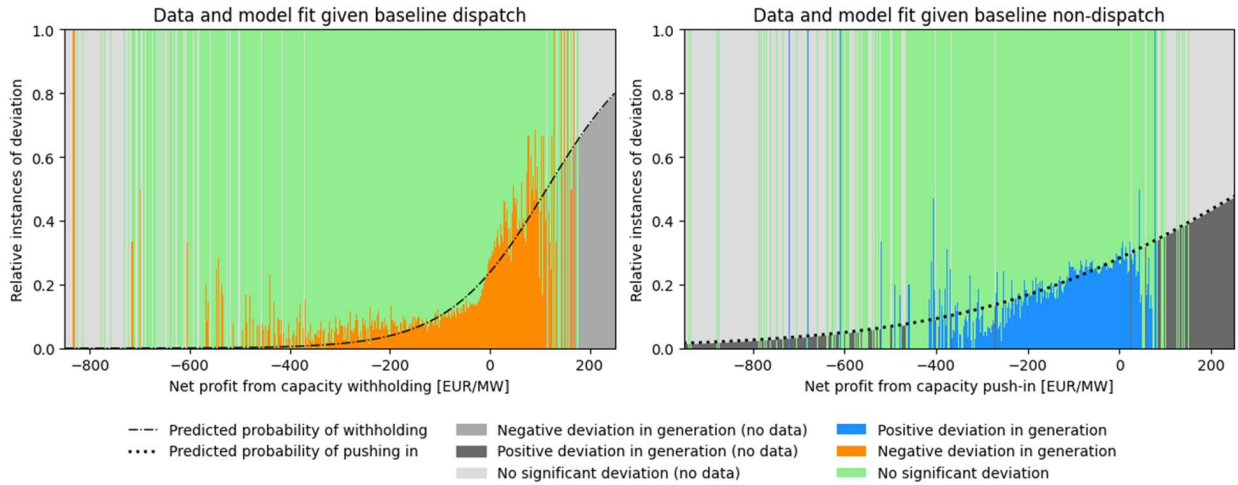

*Figure 7. Presented in two separate panels are model predicted probabilities of deviation in dispatch and observed instances of deviation by comparing empirical generation against the estimated dispatch by the competitive baseline. Suspected capacity withholding is presented on the left and suspected capacity push-in on the right. Data points of empirical observations as well as model predictions are binned over 2000 bins, along net profit, to show the average relative instances of deviation in dispatch.*

Our main finding is the following: A generation unit is more likely to be withheld if that is more profitable to do so; it is more likely to be pushed in if that is more profitable to do so. Over the six-year period and across all relevant units of coal and gas fuel types in the German wholesale electricity market, we observe two systematic patterns. First and more notably, we find a significant and positive increase in the odds of observing withholding capacity, by 1.0 % per euro increase in the net profit from inflating prices via withholding a MW capacity. This is shown by Figure 7 in the left panel. Second, the right panel shows our finding that positive deviation in dispatch is positively associated with the net profit from pushing in the respective unit, with a 0.3 % increase in the odds of observing positive deviation per euro increase in the net profit from each MW capacity push-in. Converting log odds to probability, we learn that an increase in net profit from 0 to 200 euro for withholding or pushing in 200 MW capacity[15] causes a jump in the probability of withholding from 24 % to 71 % and one in that of observing push-in from 28 % to 44 %. The colorings in Figure 7 show observed data that corroborates the model estimates in dotted lines. In short, we find that when it is more profitable to withhold, power plants are more likely to withhold and that when it is more profitable to push in, they are more likely to push in.

|  | *Slope [EUR/MW per MW change in supply]* | *Net exposure [MWh per hour]* | *Margin [EUR/MW]* | *Net profit [EUR per MW withheld]* | *Occurrence in percentile* | *Predicted probability of observing withholding* |
|---|---|---|---|---|---|---|
| *i* | 0.008 | 800 | 100 | -94 | 89 % | 11 % |
| *ii* | 0.008 | 800 | 1 | 5 | 10 % | 25 % |
| *iii* | 0.04 | 800 | 1 | 31 | 1 % | 30 % |
| *iv* | 0.04 | 4900 | 1 | 195 | 0.001 % | 70 % |

*Table 3. Model predicted probability of observing a unit being withheld, based on the net profit from withholding 1 MW under various market, company, and unit conditions in terms of slope, net exposure, and margin. Net profit is in grey because it is a calculation of the incentive composed of the three components, not new information.*

For a realistic example, the model predicts that when the net profit from withholding is negative – say at negative 94 euro when the unit withheld faces a rather high opportunity cost (indicated by its margin), with slope and net exposure at relatively favorable levels[16] in the sample (i in Table 3) – if we make a ceteris paribus change on the unit's margin from 100 to 1 EUR/MW (ii), then the net profit from withholding in that case would become positive, and correspondingly, the predicted probability of observing that unit being withheld would jump from 11 to 25 percent. If we turn to an extreme case where the slope is within top 1 percent quantile of the sample, at 0.04 EUR/MWh per MW change in supply, and net exposure at 4.9 GW for a large player in Germany, keeping margin at 1 EUR/MW (iv), the predicted probability for the unit to be withheld jumps to 70 %. Similar examples for the case of capacity push-in are included in Appendix B.

---

[15] Net profit is calculated as the euro amount net profit for each MW capacity withheld or pushed in. A 200-euro net profit can be a result of withholding or pushing in a 200 MW unit at 1 euro per MW net profit. It can also be a result of various combinations of the three components of one's economic incentive to exercise market power, which we showcase in detail hereafter.

[16] In the 6-year sample of market conditions we observe, slope at 0.008 EUR/MWh per MW capacity withheld is at about 90 percent quantile; net exposure at 800 MW is at about 80 percent quantile.

## 6.2 Robustness

Though we do not expect the fundamental incentive structure to vary over time within the sample, we conducted a robustness check on model predictions within each year in the sample period. We find that for all six years, a significant and positive association exists between potential market power abuse and net profit from deviation in dispatch. One thing worth noting is that for the year 2022, the model finds a smaller effect size for suspected withholding, compared to all other years. We suspect this results from market participants being particularly alert against legal repercussions, as there was more monitoring attention against capacity withholding during the height of the gas crisis in 2022.

|  | *2019* | *2020* | *2021* | *2022* | *2023* | *2024* |
|---|---|---|---|---|---|---|
| Net profit from capacity withholding [EUR/MW] | 0.0408*** | 0.0421*** | 0.0125*** | 0.0081*** | 0.0154*** | 0.0127*** |
| Net profit from capacity push-in [EUR/MW] | 0.0068*** | 0.0079*** | 0.0038*** | 0.0053*** | 0.0053*** | 0.0050*** |

Table 4. Logit regression results for year-specific models respectively for suspected withholding among negative deviation in dispatch and for suspected push-in among positive deviation in dispatch. Note: * signifies statistical significance at p-value of 0.05, ** at p-value of 0.01, *** at p-value of or below 0.001.

We also ran technology-specific models as a robustness check on our main results. The results show significant profit-driven deviation in dispatch across production types: we find suspected withholding across the board and suspected push-in in gas-fired units. The details on these results are included in Appendix B.

From company-specific model runs, we find that 10 out of 15 companies in the sample show both suspected capacity withholding and push-in that follow their hourly economic incentives. The remaining 5 companies show significant associations that align with our hypothesis with one of the two types of dispatch deviations. Figure 8 presents the results of company-specific logit models, ranking larger to smaller companies from left to right, by total MW generation capacity.

The heterogeneity in company model results can be explained by company differences in their propensity to respond to potential net profits of market power abuse. This includes their heterogenous risk attitudes in front of monitoring agencies, which we do not explicitly study in this analysis.

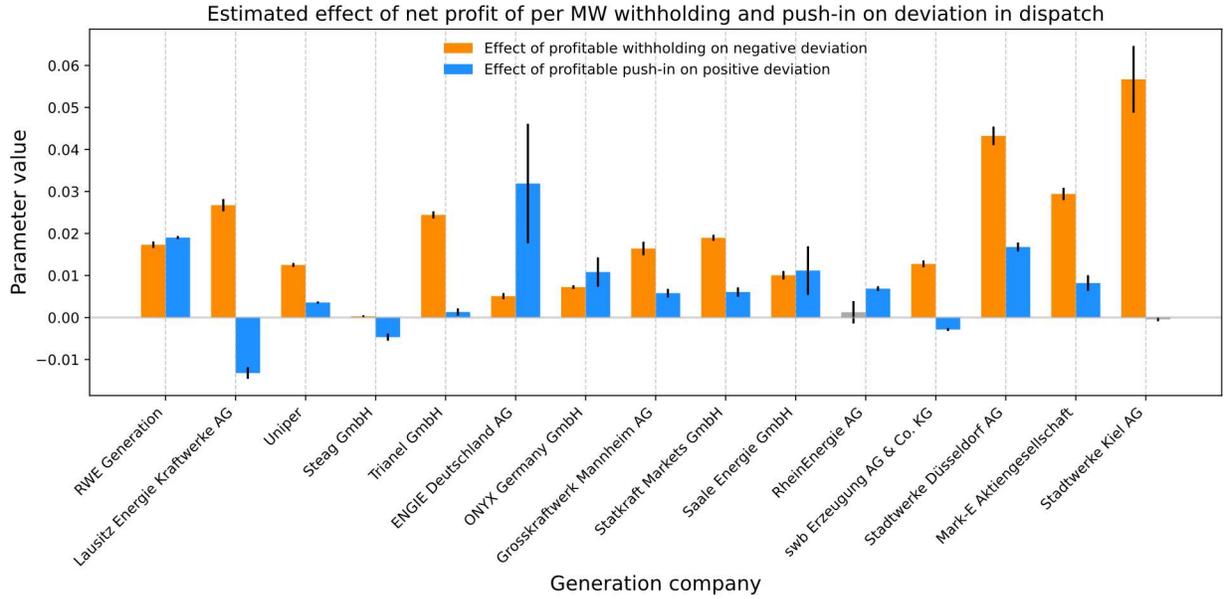

*Figure 8. Log odds ratio parameter estimates across company-specific model results. Black lines indicate confidence intervals of the estimated effects of profitability of market power abuse on deviation in dispatch, in respective directions.*

## 6.3 Sensitivities

When assuming a lower hedge rate, the model estimates more generation output a company has selling on the wholesale market, thus also a higher incentive to withhold and a lower incentive to push in. When the hedge rate assumption does not match reality, the model then overestimates the net profit from withholding and underestimates that from push-in. Consequently, we observe lower coefficient estimates for withholding and higher coefficient estimates for push-in (see Table 5).

The fact that the model finds a positive and larger effect from the net profit from push-in on deviation in dispatch, even when hedge rate is assumed to be zero and the economic incentive for push-in assumed to be non-existent, is evidence consistent with public information that the hedge rate is in reality not zero.[17]

| *Hedge rate* | *Effect of net profit of withholding on negative deviation in dispatch* | *Effect of net profit of push-in on positive deviation in dispatch* |
|---|---|---|
| **100 %** | 0.0104*** | 0.0034*** |
| **70 %** | 0.0096*** | 0.0036*** |
| **0 %** | 0.0067*** | 0.0040*** |

*Table 5. Model results with varying levels of hedge rate assumptions. Note: \* signifies statistical significance at p-value of 0.05, \*\* at p-value of 0.01, \*\*\* at p-value of or below 0.001.*

---

[17] Generation companies periodically disclose their hedge ratios for thermal units. For example, in an investor presentation in 2020, RWE Generation has reported to have already hedged above 90 percent of its thermal production from 2020 to 2023.

The main model specification focuses on the marginal effects of net profit from market power abuse, per MW capacity withheld or pushed in. As we discretize deviation in dispatch, this seemingly constitutes a mismatch between the outcome variable, deviation in dispatch status, and the predictor variable, net profit from deviation in per-MW generation. By focusing on withholding from minimum load to zero generation and push-in from zero to minimum load, the binary measurement of deviation in dispatch status is in effect the marginal change in generation from and to zero output, similar to marginal changes in net profit, due to the fact that generation units face unit-commitment constraints and cannot steadily run below their minimal load requirements.[18]

Meanwhile, the effect of net profit from deviation on generation deviation can be non-linear, along varying levels of capacity deviated. Therefore, we check if the net profits specific to a 1/3-unit commitment also have similar association with suspected uncompetitive generation patterns. The regression results are presented in Table 6. Consistent with main model results, we find positive associations between the net profits from market power abuse and the deviations in dispatch in both directions. The effect sizes, however, are smaller because net profits now are overall much larger in value, whose per-unit increase in euro involves multiple MW capacity.

|  | *Negative deviation in dispatch (suspected capacity withholding)* | *Positive deviation in dispatch (suspected capacity push-in)* |
|---|---|---|
| *Intercept* | -1.20*** | -0.93*** |
| *Net profit of withholding 1/3-unit capacity [EUR]* | $4.39 \times 10^{-5}$*** |  |
| *Net profit of pushing in 1/3-unit capacity [EUR]* |  | $2.77 \times 10^{-5}$*** |
| *$R^2$ (McFadden)* | 2.7 % | 0.7 % |
| *Number of observations* | 727741 | 686341 |

*Table 6. Logit regression results for two models: one for suspected capacity withholding among negative deviation in dispatch, the other for suspected capacity push-in among positive deviation in dispatch. Note: \* signifies statistical significance at p-value of 0.05, \*\* at p-value of 0.01, \*\*\* at p-value of or below 0.001.*

## 7. Discussion

The 2011-introduced REMIT Article 5 prohibits market manipulation of various kinds on wholesale energy markets. The cases of REMIT breaches published by the EU Agency for the Cooperation of Energy Regulators (ACER 2023) so far have found market manipulation via issuing non-genuine trade orders to mislead the market (ANRE 2022) and artificially creating system imbalance (Ofgem 2020a) and cross-border price difference (Forsyningstilsynet 2025). One case pertaining capacity withholding was found in UK, where generation company InterGen falsely declared unavailability during peak winter hours in 2016, creating an appearance of shortage that allowed it to profit from National Grid's higher spending on the balancing mechanisms (Ofgem 2020b). Another case was found in France, where the generation company EDF omitted the unavailability of its nuclear fleet when it was in fact experiencing outage (Journal officiel de la République française 2022). We suspect that more cases of market power abuse deserve monitoring attention and hope that our analysis provides the tool needed to spark investigations.

---

[18] The minimal load requirement usually ranges from 20 to 40 percent of a generation unit's installed capacity, depending on production type and unit characteristics. In this sensitivity run, we take the one-third of unit installed capacity as the MW capacity required for any deviation in dispatch.

The results of our analysis show systematic profit-driven deviation in dispatch for coal- and gas-fired generation units in the German wholesale electricity market. On average, the expected volume of generation withheld and pushed in from the 40 turbines in our sample take up about 4 % of the total annual load per year; the associated change in the hourly spot price averages at 4 EUR/MWh and goes up to 160 EUR/MWh inflated due to withholding and due to push-in, it averages at a 3 EUR/MWh reduction and dives down by a maximum of 38 EUR/MWh (see Table B2). Though results are contingent on the assumptions we take to set up the analysis, they nevertheless provide compelling evidence of systematic market power abuse that warrants monitoring attention.

We choose to be conservative in our methods whenever we face sufficient uncertainty in our approach, to minimize the risk of mistaking competitive generation behavior for market power abuse. This is most salient in three places: First, we simulate competitive, generation-unit-specific dispatch over a thousand iterations of varying assumptions about the unit's thermal efficiency, local fuel cost, and start-up cost. We disregard unit-hours where above 5 % of the simulation results do not agree. Second, we disregard market power abuse via reporting technical outage and treat all reported unavailable capacity to be truly unavailable, not potentially withheld. This is because the quantification of strategic decision-making in outage reporting has large uncertainty and high heterogeneity at the generation unit level. Third, we discretize generation to dispatch status, as a continuous measurement comes with high uncertainty in model prediction. This limits our ability to detect partial capacity withholding of units that are not turned off and partial capacity push-in of units that are already running to some extent. Ultimately, taking the conservative route when we face high model uncertainty allows us to be confident in the model results we find significant within the limited scope: they give a lower bound of suspected market power abuse. Hereafter, we discuss further considerations in discretizing generation and treating outage report data, as well as additional limitations of this analysis.

We discretize deviation in generation to binary dispatch and non-dispatch mainly because generation decisions are ultimately binary – it is simply not optimal to run a power plant halfway between minimum and maximum generation capacity within the constraints, either in terms of strategic gains or power plant operation. Meanwhile, power plants sometimes run under less optimal decision-making, a continuous measure of deviation in generation can therefore better the accuracy of model predictions, though at the cost of larger uncertainty around the predictions. We believe the benefit of increasing complexity of the model is minimal and does not outweigh the reduced confidence in model results.

Unit reported outage and unavailability are taken at face value in this analysis because we are not entirely certain about the extent of strategy disguised in technical reports by generation unit owners. As previous literature concludes that there is capacity withholding in reported outage, assuming no withholding under reported outage gives a downward bias in our estimates of suspected withholding. Nevertheless, we still find a significant positive association between our rather conservatively estimated deviation in dispatch and net profit from market power abuse.

An important limitation of this analysis is that our estimate of deviation in dispatch does not consider generation called on for redispatch, due to the lack of publicly available data. Future research is needed for a more extensive picture of deviation in unit generation.

Company's revenues and costs on related markets are not considered in this analysis, such as balancing and ancillary services. For instance, our results can be biased by generation decisions made due to profits on the balancing market instead of the day-ahead one, though previous research mentions that in Germany, considerations for profiting from the balancing market are generally secondary to those from the day-ahead market.

Renewable technologies are not studied in this analysis, but that does not mean they do not possess market power or cannot exercise it. We take wind and solar power generation as input for our analysis for coal- and gas-fired units but do not study their generation behavior mainly because there is a lack of high-granularity data that allows for the mapping of unit generation to company. Additionally, VRE have faced different incentive structures than thermal technologies, where they are almost never exposed to the spot market. In the sample period in Germany, for example, most solar and wind power generation is subsidized under the Erneuerbare-Energien-Gesetz (EEG) and receive a feed-in tariff or market premium instead of the market price, regardless of their generation in quantity. With that said, VRE units can still be used as a tool for price manipulation, if the owner potentially benefits from its other units that are exposed to spot prices. In this case, withholding via curtailing VRE attracts less monitoring attention, though it is less optimal than withholding peaking units due to VRE's larger opportunity costs. We also expect capacity push-in via VRE to be less effective, as renewable ownership is diverse and more price-elastic. The lowered price via push-in can trigger economic curtailment from the rest of the VRE supply when it exceeds the threshold where the subsidy no longer covers the loss of further injection. We recommend future research on the incentive structures of VRE units to exercise their market power, as they become increasingly influential in recent and future power markets.

The introduction of flow-based market coupling and increase in interconnector capacity likely have helped reduce the concern for market power abuse. We do not explicitly model the effect cross-border trade has had on market power abuse, though the result of such effect is included in our measurement of the slope of the supply curve. Future research is needed to measure the potentially neutralizing effect of interconnection on market power.

## 8. Conclusion

Market power abuse is typically difficult to pin down in electricity markets, due to its volatile nature at the hourly level. We provide a measure of the hourly net profit from exercising one's market power and explain unit deviation in dispatch with variations in price-altering incentives between hours. For thermal, dispatchable generation units in the German wholesale market, we find, across the board, significant associations between a unit's net profit from potential price-influencing withholding and push-in of capacity and observed deviation in dispatch. We interpret this as empirical evidence of market power abuse, where a company's economic incentive significantly influences its unit generation decisions, even and perhaps especially when it harms market functioning.

Our results have important policy implications for industrial organization and sequential market design in electricity markets. We suggest that to minimize the opportunity of market power abuse, company size needs to be closely monitored, and company hedge profiles should be considered by monitoring agencies. More importantly, the forwards market presents an immense potential to neutralize the incentives to exercise market power, while currently, it is not set up for that. Economic incentives for companies to depress market prices emerge partly due to the common base futures structure of product offering on the forwards market. We suggest offering products of higher granularity than the common on-peak/off-peak base futures, to enable the market to automatically align company hedging with mitigation of market power abuse. Forwards markets therefore have the potential to fundamentally neutralize economic incentives to exercise one's market power, if hourly future products are made available. Granted, companies will still face an inevitable trade-off between the benefit of hedging against price risks and the loss in premium by mitigating their own market power.

**CRediT authorship contribution statement**

Alice Lixuan Xu: Writing – original draft, Visualization, Software, Project administration, Formal analysis, Data curation, Conceptualization.

Jorge Sánchez Canales: Writing – original draft, Visualization, Software, Formal analysis, Data curation, Conceptualization.

Chiara Fusar Bassini: Writing – review & editing, Visualization, Data curation, Conceptualization.

Lynn H. Kaack: Writing – review & editing, Visualization, Supervision, Funding acquisition, Conceptualization.

Lion Hirth: Writing – review & editing, Visualization, Supervision, Funding acquisition, Conceptualization.

**Data availability**

The code and data needed to replicate the analysis are available under an open license and can be found at https://github.com/alicelixuan/MPA-electricity.

**Acknowledgements**

This work is supported by the German Federal Ministry of Education and Research (BMBF) via the ARIADNE Project (FKZ 03SFK5K0-2) and the ML-Strom Project (FKZ 16DKWN102) and is part of the German Recovery and Resilience Plan (DARP), financed by NextGenerationEU, the European Union's Recovery and Resilience Facility (ARF).

## 10. Appendix A: Optimization model for competitive power plant dispatch

This appendix describes the optimization model we use to predict the competitive dispatch behavior of thermal power plants. The model operates over a planning horizon of $H$ hourly time steps and focuses on a single generation unit. Its objective is to maximize profit, defined as the revenue from electricity sales minus operational costs, while respecting its operational constraints. The model focuses on a single unit at a time, based on the assumption that in a perfectly competitive market, each unit operates independently, without considering interactions with other units.

The electricity prices are treated as being exogenous to the dispatch of the units and are as realized, meaning they are taken as given based on the historical series of the German day-ahead market. Therefore, the optimization model assumes perfect foresight of electricity prices over the planning horizon.

We built the model using Pyomo, an optimization modeling framework in Python (Bynum et al. 2021) and used the GNU Linear Programming Kit (Makhorin 2012) as the solver. The optimization model is solved iteratively for each unit $i$ over planning horizons $m$ of length $H_m$, with a one-day overlap to mitigate boundary effects. To account for parameter uncertainty, a Monte Carlo simulation approach is applied.

*Model formulation*

The optimization model is formulated as a Mixed-Integer Linear Program (MILP) using Pyomo's ConcreteModel class. This formulation corresponds to one run of the model, meaning that it applies for one unit $i$ and one planning horizon $k$, whereas our whole sample is made of 40 units and up to 74 planning horizons for each unit. The model incorporates the following components:[19]

Sets

- $H_m$: The set of time periods $h$ in the planning horizon $m$.

Parameters

- $K_i$: Installed nameplate capacity of the unit, in MW.
- $G_i^{min}$: Minimum dispatchable power output of the unit, in MW (Schill, Pahle, and Gambardella 2017, 8).
- $p_h$: The electricity wholesale market price at time $h$, in €/MWh.
- $c_{i,h}^{var}$: The variable cost of generating electricity at time $h$, in €/MWh electric. This consists of fuel (in €/MWh thermal) and carbon costs (in €/MWh thermal), divided by plant thermal conversion efficiency $\sigma_i$, such that:

---

[19] For a general introduction to what these components are in optimization modeling, see (Bynum et al. 2021, 38).

$$c_{i,h}^{var} = \frac{c_{i,h}^{fuel} + c_h^{carbon}}{\sigma_i}$$

- $c_{i,h}^{start}$: Fixed startup cost at time $h$, in €. Based on functional and numeric estimates from Schill, Pahle, and Gambardella (2017), total startup costs are the sum of startup depreciation costs (in €/MW) and cold startup fuel costs (in €/MW) times the installed capacity. The cold startup fuel requirements are derived from a cold startup fuel requirement ($Q_t$, in MWh thermal/MW), a cold startup factor ($r_t$) and the cost of the fuel, including carbon.

$$c_{i,h}^{start} = K_i * [c_i^{depreciation} + Q_t * r_t * (c_{i,h}^{fuel} + c_h^{carbon})]$$

Decision variables

- $G_{i,h}$: Power generated at time $h$, in MW. $G_{i,h} \in \{0\} \cup [G_i^{min}, K_i]$
- $\hat{d}_{i,h}$: Binary variable indicating whether the plant is on ($\hat{d}_{i,h} = 1$) or off ($\hat{d}_{i,h} = 0$) at time $h$. $\hat{d}_{i,h} \in \{0, 1\}$.
- $s_{i,h}$: Binary variable indicating whether the plant starts up ($s_{i,h} = 1$) or not ($s_{i,h} = 0$) at time $h$. $s_{i,h} \in \{0, 1\}$.

Objective function

The objective is to maximize the total profit over the planning horizon $H$:

$$Maximize \ \pi_{i,H} = \sum_{h \in H} (p_h * G_{i,h} - c_{i,h}^{var} * G_{i,h} - c_{i,h}^{start} * s_{i,h})$$

Where:

- $p_h * G_{i,h}$: Revenue from selling electricity.
- $c_{i,h}^{var} * G_{i,h}$: Costs of generating electricity (fuel and carbon).
- $c_{i,h}^{start} * s_{i,h}$: Startup costs.

Constraints

The model enforces the following operational constraints:

- Maximum dispatch: The power output during each time period can only be non-zero if the plant is on and is upward bounded by the installed nameplate capacity.

$$G_{i,h} \leq K \cdot \hat{d}_{i,h} \qquad \forall h \in H$$

- Minimum dispatch: The plant must generate at least the minimum dispatchable power $G_i^{min}$ when it is on:

$$G_i^{min} \cdot \hat{d}_{i,h} \leq G_{i,h} \qquad \forall h \in H$$

- State: The plant can only be on if it generates electricity, or the plant must be off if the output is 0.

$$\hat{d}_{i,h} \leq G_{i,h} \qquad \forall h \in H$$

- Startup: The plant is considered to have started up if it transitions from off ($\hat{d}_{i,h} = 0$) to on ($\hat{d}_{i,h} = 1$):

$$s_{i,h} \geq \hat{d}_{i,h} - \hat{d}_{i,h-1} \qquad \forall H \in H, \quad h > 1$$

*Solution approach*

The optimization model is solved iteratively for each unit in the dataset using the GLPK solver (General Linear Programming Kit). To manage computational complexity and reflect realistic decision-making horizons, the time series data is divided into planning horizons $m$ of equal length. For each unit $i$ and planning horizon $m$, the model determines the optimal dispatch and startup decisions for each time period $h$. The results are then aggregated across planning horizons.

Planning horizon

We denote the length of the planning horizon $m$ as $H_m$, which is set to encompass one month (30 days), corresponding to 720 hours. To mitigate corner solutions,[20] we extend the optimization to include an additional day, resulting in $H_m = 744$ hours for all planning horizons (except the last one, whose length is determined by the number of observations left). The solution of the last 24 time periods $h$ corresponding to the additional day is calculated twice, as the last day of one planning horizon $m$ and the first day of the next one $m + 1$.

$$H_m = 744 \qquad \forall m \neq M,$$

$$H_M \leq 744$$

For all planning horizons except the first, the model uses the solution from the end of the previous horizon $m - 1$ as the initial condition for the current horizon $m$. Specifically, the generation $G_{i,h}$, startup $s_{i,h}$, and on/off state $\hat{d}_{i,h}$ from the first hour of the last day ($h = H_{m-1} - 23$) are carried over to $h = 1$ in the current horizon $m$, while the solutions of the remaining 23 hours of the last day are discarded due to the one-day overlap and recalculated in the current time horizon $m$. With this approach we ensure continuity between planning horizons and avoid unrealistic transitions in plant operations.

---

[20] For example, the model might shut the plant off in the last hour due to negative prices, even though positive prices in the subsequent hour could justify keeping it running.

Monte Carlo Simulations

To account for uncertainties in input parameters, we apply a Monte Carlo simulation. In each iteration, we sample independent multipliers for fuel costs $c_{i,h}^{fuel}$, thermal conversion efficiency $\sigma_i$ and cold start factor $r_t$ from a normal distribution with mean 1 and standard deviation 0.05:

$$m_i^{\sigma}, m_i^{c^{fuel}}, m_i^{r} \sim N(1, 0.05)$$

In other words, each iteration is run with modified unit-specific fuel costs and thermal conversion efficiency using a normal distribution that captures a ±10% variation around the mean within a 95% confidence interval.

These variations propagate through the model, influencing both the variable costs of generation $c_{i,h}^{var}$ and startup costs $c_{i,h}^{start}$ as described above, both of which directly affect the objective function. The simulation is run 1000 times to generate a range of possible scenarios, capturing the impact of parameter uncertainty on dispatch decisions and profitability. This approach allows us to obtain a measure of our overall confidence on our dispatch prediction, as we are able to differentiate between hours that are always predicted to be dispatching across model runs from hours that change status depending on the assumptions on parameters.

*Model outputs*

The model produces the optimal dispatch schedule for each generation unit over the planning horizon. This schedule includes the unit's power output and operational states (on/off) for each time period. While the continuous dispatch decisions $G_{i,h}$ are crucial for assessing the unit's operational profitability, the primary output of interest is the binary state variable ($\hat{d}_{i,h}$), which indicates whether the plant is running at time $h$.

To incorporate uncertainty, the Monte Carlo simulation generates multiple realizations of the optimal state variable across iterations. The final output is the **average state** ($\bar{d}_{i,h}$), defined as the fraction of Monte Carlo iterations in which the unit was predicted to be operational during each time period $h$:

$$\bar{d}_{i,h} = \frac{1}{N} \sum_{n=1}^{N} \hat{d}_{i,h}^{n}, \qquad \bar{d}_{i,h} \in [0, 1]$$

where:

- $N$ is the total number of Monte Carlo iterations, and
- $\hat{d}_{i,h}^{n}$ is the state of the unit in time period $t$ during iteration $n$.

The average state provides a probabilistic measure of the unit's dispatch behavior, reflecting its likelihood of operation under varying input conditions.

*Limitations*

- Time-invariant efficiency and startup costs: The model assumes that thermal efficiency is fixed for each Monte Carlo iteration and does not vary with temperature or output level. In reality, efficiency changes dynamically with operating conditions, but incorporating this would require making efficiency a decision variable, resulting in a non-linear model. Similarly, startup costs depend on down-time, as the longer a plant is turned off, the costlier it is to turn it up (Schill, Pahle, and Gambardella 2017).
- Lack of ramping constraints: The model assumes full flexibility beyond startup costs, meaning output can be freely adjusted between minimum and maximum capacity at each time step, as long as the plant is online. This might ignore real-world technical limitations on how quickly a plant can ramp up or down.
- Perfect foresight of power prices: The optimization is performed using realized prices for a full planning horizon of one month plus one day ($H = (30 + 1) * 24 = 744$). In practice, power plants make decisions in a day-ahead market with price uncertainty. This assumption particularly affects startup costs, as real-world operators may act more conservatively due to break-even concerns in the face of fundamental price risk.
- No planned maintenance or unplanned outages: The model does not dynamically account for plant maintenance or forced outages. Instead, these factors are addressed at a later stage when constructing the deviation variable for the econometric model. This is not so problematic for forced outages (as they are themselves unforeseeable), but planned maintenance could affect the optimal solution when determining startup cycles.
- No market clearing mechanism: The model does not ensure supply and demand balance, as it does not include a market-clearing process. While this is sufficient for our purposes, it introduces uncertainty in dispatch decisions near the break-even point, where real-world power exchanges might favor smaller (but more expensive) units to better match market-clearing conditions.

*Summary statistics of input parameters and output variables*

|  | Notation | Unit | Mean | Median | Min | Max | Domain |
|---|---|---|---|---|---|---|---|
| Input parameters | | | | | | | |
| Price | $p_h$ | €/MWh | 84.2 | 49.2 | -500.0 | 871.0 | R |
| Variable cost | $c_{i,h}^{var}$ | €/MWh | 91.4 | 55.3 | 18.5 | 1055.8 | Positive R |
| Startup cost | $c_{i,h}^{start}$ | 1000 € | 73.7 | 55.7 | 2.4 | 896.5 | Positive R |
| Installed capacity | $K_i$ | MW | 474.0 | 321.0 | 112.0 | 1060.0 | Positive R |
| Minimum dispatch | $G_i^{min}$ | MW | 175.7 | 172.4 | 17.0 | 424.0 | Positive R |
| Decision variables | | | | | | | |
| Dispatch | $G_{i,h}$ | MW | 257.5 | 5.6 | 0 | 1060 | 0 or between $G_i^{min}$ and $K_i$ |
| State | $d_{i,h}$ | - | 0.47 | 0.25 | 0 | 1 | {0, 1} |
| Startup | $s_{i,h}$ | - | 0.0043 | 0 | 0 | 1 | {0, 1} |

## 11. Appendix B: Additional figures and tables

*Piecewise linear estimates for the slope of the supply curve, within each regime*

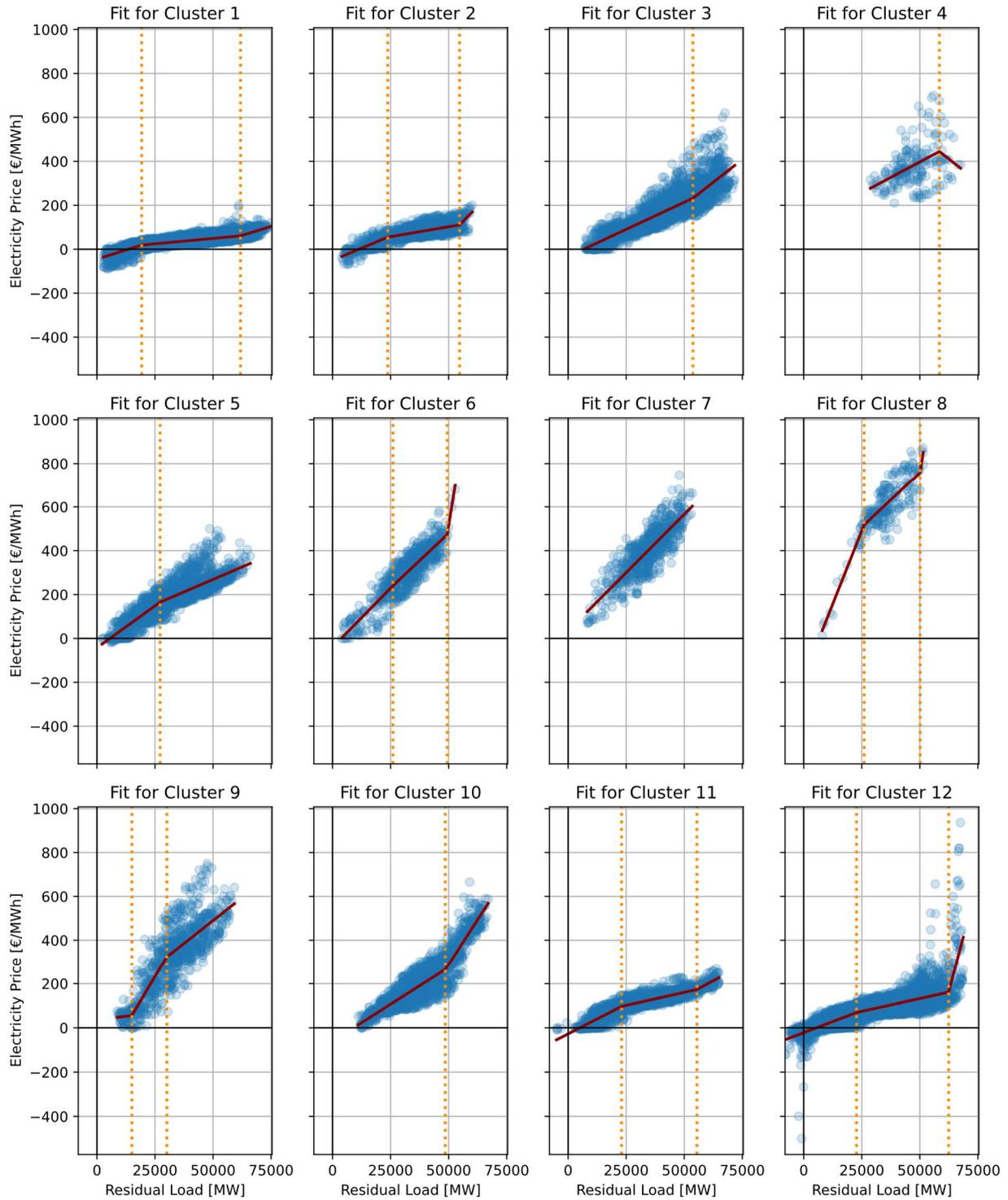

*Figure B1. Piece-wise linear fits within all 12 clusters, representing distinct regimes of electricity market supply for the sample period 2019-2024.*

*Tangible interpretation for the case of capacity push-in*

|     | Slope (EUR/MW per MW change in supply) | Net exposure (MW) | Margin (EUR/MW) | Net profit (EUR per MW withheld) | Occurrence in percentile | Predicted probability of observing withholding |
|-----|---|---|---|---|---|---|
| i   | 0.006 | -800  | -100 | -95 | 82 %     | 22 % |
| ii  | 0.006 | -800  | -10  | -5  | 9 %      | 28 % |
| iii | 0.02  | -800  | -5   | 11  | 1 %      | 29 % |
| iv  | 0.04  | -9000 | -5   | 175 | 0.0001 % | 42 % |

Table B1. Model predicted probability of observing a unit being pushed in, based on the net profit from pushing in 1 MW under various market, company, and unit conditions in terms of slope, net exposure, and margin. Net profit is in grey because it is a calculation of the incentive composed of the three components, not new information.

*Expected volume withheld and pushed-in, based on market-level model prediction*

|  | Expected volume withheld, relative to load | Average expected change in spot price, associated with withholding [EUR/MW] | Max. expected change in spot price, associated with withholding [EUR/MW] | Expected volume pushed in, relative to load | Average expected change in spot price, associated with push-in [EUR/MW] | Min. expected change in spot price, associated with push-in [EUR/MW] |
|---|---|---|---|---|---|---|
| Six-year sample | 2.4 % | 3.5 | 160.4 | 1.8 % | -2.9 | -37.8 |
| 2019 | 2.8 % | 2.0 | 11.2  | 1.5 % | -1.6 | -14.9 |
| 2020 | 2.4 % | 1.8 | 13.0  | 2.2 % | -2.3 | -14.9 |
| 2021 | 2.2 % | 3.7 | 20.6  | 1.2 % | -2.1 | -15.6 |
| 2022 | 1.3 % | 5.7 | 59.7  | 1.2 % | -4.8 | -37.8 |
| 2023 | 1.7 % | 3.9 | 139.2 | 1.7 % | -3.3 | -19.8 |
| 2024 | 1.7 % | 4.0 | 160.4 | 1.9 % | -3.4 | -17.5 |

Table B2. Expected capacity withholding and push-in, based on model predictions. Shown here are annual total expected volume being withheld (on the left) and pushed in (on the right), as well as the associated change in hourly spot prices.

*Technology-specific model runs*

Table B3 shows deviation in dispatch for units of different production types, as an output of the competitive dispatch model in the first step of our method. We find deviation in dispatch to have a strong heterogeneity among different technologies. Combined cycle gas turbine (CCGT) units stand out to have on average more negative deviation from the competitive benchmark, indicating possible capacity withholding. On the other hand, lignite and other types of gas (non-CCGT) units have more positive deviation in dispatch, as compared to the competitive benchmark, consistent with possible capacity push-in.

| *Production type* | *Negative deviation* | *No significant deviation* | *Positive deviation* |
|---|---|---|---|
| **All units pooled** | 8.0 % | 82.5 % | 9.5 % |
| *Of all units:* | | | |
| *Lignite* | 0.8 % | 86.3 % | 12.9 % |
| *Hard coal* | 7.3 % | 84.2 % | 8.5 % |
| *CCGT* | 17.7 % | 79.1 % | 3.2 % |
| *Gas other types* | 2.0 % | 81.9 % | 16.1 % |

*Table B3. Frequency of deviation in dispatch for all units in the analysis, summarized by production types.*

Figure B2 summarizes logit results from technology-specific models, where we run the same regression model respectively for units of various fuel types. We find that for three of the four production types in the sample, deviation in dispatch has a strong positive association with the net profit from potential capacity withholding, though the case for push-in is less clear. We suspect that unit operation pertaining to system balancing plays a substantial role for the generation behavior of coal units.

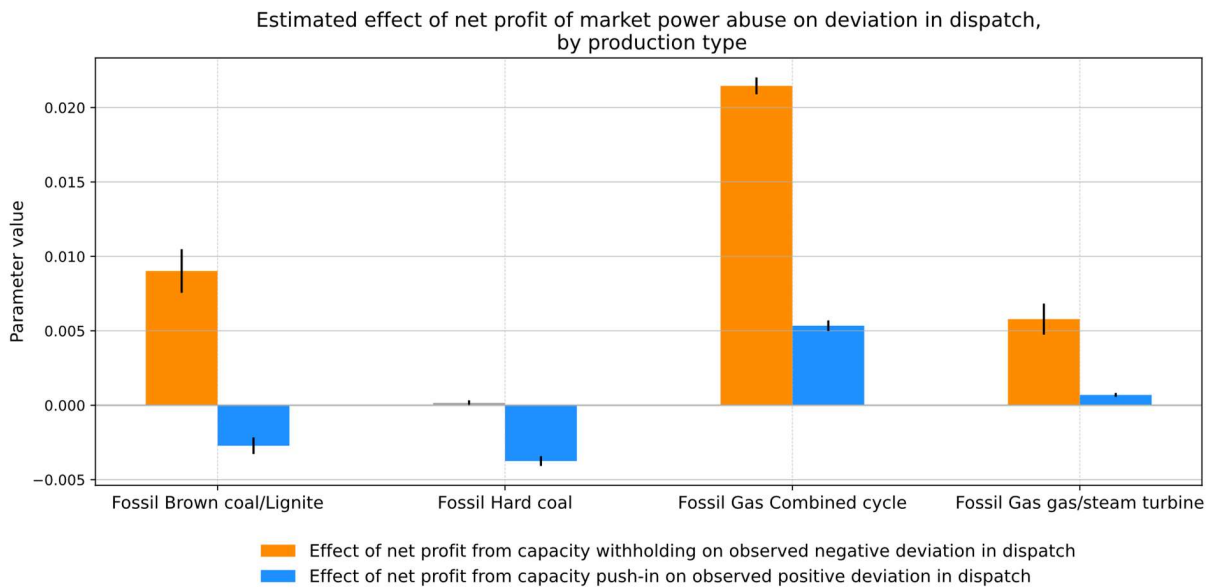

*Figure B2. Results of technology-specific logit regression models, where coefficients of the variable – net profit from market power abuse – is presented. Each bar shows individual model estimates of the marginal change in the log odds of seeing a unit being withheld or pushed in, following 1 euro increase in the net profit from withholding or pushing that unit into supply. The confidence intervals at 95% are shown by the black lines on top.*